# Review Text as a Leading Indicator of Displayed Reputation in Platform Rating Systems: Evidence from 34 U.S. Short-Term Rental Markets

**Ali Safari**
Department of Information Technology and Decision Sciences, University of North Texas
ali.safari@unt.edu

## Abstract

Rating systems on accommodation platforms suffer from a familiar problem: nearly every listing displays a nearly perfect score, so the number that is supposed to separate good listings from bad ones barely varies. Whether the review text accumulating beneath those scores still carries usable information is an open question. I ask a dynamic version of it: does the text guests have already written predict where a listing's displayed rating moves next? Treating text and ratings as parallel channels that aggregate guest experience at different speeds, I construct a prespecified sentiment index from the complete review history of each listing in a two-wave panel of more than two hundred thousand listings across 34 U.S. markets. Because the broader project had explored these data before, I locked the model and its falsification checks in advance and reserved half of the markets, untouched, for a single confirmatory estimation. On those held-out markets, warmer past text predicts a small but precisely estimated upward movement of the displayed rating over the following year. Listings that received no new reviews show no such movement, the association survives host fixed effects, and no single market drives it. The results indicate that the platform's rating aggregation discards information its own review text retains. I discuss what this leading-indicator property implies for the design of reputation displays. The text index is a defined dictionary-based instrument that has not been validated against human judgment, and I state that boundary plainly.

**Keywords:** reputation systems, online reviews, text mining, rating dynamics, sharing economy, panel data

## 1. Introduction

A typical Airbnb listing in my data displays an overall rating between 4.5 and 5.0 stars, and it acquired that rating years ago. Zervas et al. (2021) documented the pattern across nearly the entire platform: where every stay is above average, the displayed score stops separating good listings from bad ones. Bridges and Vásquez (2018) asked the natural question of whether near-universal positivity makes reviews meaningless. In an earlier version of this work, I studied that saturation across six U.S. regions and reached the same descriptive conclusion. That result raised a question the earlier version could not answer: if the number on the page has stopped moving, does anything in the review stream still move ahead of it?

This paper answers with a design change rather than a bigger model. The saturation literature treats the displayed rating as the endpoint of reputation, and asks whether reviews still influence prices or demand given that endpoint. I challenge the assumption built into that framing, the assumption that the displayed rating and the review text carry the same information on different scales. If text and score were redundant channels, the text accumulated before a snapshot would tell me nothing about where the score moves next. If instead guests write information into text that the coarse rating integrates only slowly, then yesterday's text should predict tomorrow's rating change. That is a falsifiable, directional claim about the platform's own signal, and it needs no price data, no demand model, and no assumption about guest behavior.

I test the claim on a two-wave panel built from 13.75 million reviews across 34 U.S. markets, with 216,248 listings matched across the 2025 and 2026 snapshots. From each listing's complete pre-2025 review history I construct an aspect-based text index. A prespecified dictionary assigns sentences to accuracy, cleanliness, check-in, or communication; the VADER tool scores their sentiment; listing means shrink toward market priors; and the four standardized aspect signals average into one index. The outcome is the movement of the displayed overall rating between the two snapshots.

Because I explored candidate constructs on these data before settling on this question, I treated my own estimates with suspicion. I locked the research question, the model, the falsification battery, and the success criteria in a written prespecification, split the 34 markets by seeded randomization into 17 discovery and 17 confirmation markets, and estimated the locked model exactly once on the confirmation half. On those markets, a standard deviation of warmer text predicts an upward movement of the displayed rating of about a hundredth of a star over the following year, an estimate that is small, precise, and stable. A placebo group of listings with no new reviews shows no rating movement. The estimate survives host fixed effects among multi-listing hosts and stays positive when I drop any one of the 17 confirmation markets. Section 5 reports the full estimates.

The contribution is a documented empirical regularity for the reputation-systems literature: cumulative review text is a leading indicator of the platform's own displayed rating. Its interest lies in the direction and the design, not the magnitude. Text written before the snapshot predicts movement in a score that already summarizes that same history, which means the rating aggregation discards information that the text retains. For platform designers, the finding identifies untapped signal inside data the platform already owns. For researchers, it reframes text-versus-rating comparisons from a question of consistency at a point in time to a question of timing.

I am equally explicit about what this paper does not claim. The design is associational; new-guest selection and persistent latent quality are live alternative pathways, and I discuss both. The text index is a defined dictionary-plus-VADER instrument that has not been validated against human judgment, a boundary I return to in the limitations. And the signal is general rather than sharply aspect-specific: it predicts overall rating movement, and I do not claim that mismatches between a specific aspect's text and its rating carry the effect.

## 2. Background

### 2.1 Saturated ratings and the trust problem

Reputation mechanisms are the trust infrastructure of peer-to-peer exchange (Ert and Fleischer 2019). On accommodation platforms the mechanism has a well-documented pathology: scores cluster so tightly near the ceiling that they lose discriminating power. Zervas et al. (2021) compared Airbnb ratings with ratings of the same properties elsewhere and found the platform's distribution strikingly compressed at the top. Bridges and Vásquez (2018) reached the same point from a discourse perspective. The compression has identifiable sources in the reviewing process itself. Fradkin, Grewal, and Holtz (2021) showed experimentally that when feedback was hidden until both parties reviewed, retaliation and reciprocation fell and ratings dropped, which identifies strategic reviewing as one inflation channel. Abrardi et al. (2026) added that personal interaction between hosts and guests biases the numeric rating upward. Design interventions move the channels differently: inducing more reviews changes market outcomes only modestly (Fradkin and Holtz 2023), while the simultaneous-reveal policy changed the informational content of review text itself (Mousavi and Zhao 2022). Consistent with saturation, Yang (2026) found that review volume carries independent signal value mainly at the low end of the revenue distribution, where information is scarce. A compressed score does not mean guests learned nothing; it means the display integrates what they learned slowly and coarsely.

## 2.2 Text and ratings as parallel channels

A second stream compares the two channels directly. Pocchiari, Proserpio, and Dover (2025) synthesized the online-review literature across the stages of review creation, engagement, impact, and insight, and the relation between text and numeric evaluation runs through all four stages. Mudambi and Schuff (2010) established that review text carries decision-relevant information beyond the star rating. Almansour et al. (2022) reviewed the discrepancy literature and concluded that text and score frequently disagree, enough that score-trained sentiment labels inherit real noise. Cheng and Jin (2019) showed that Airbnb review text is organized around a small set of recurring concerns, location, hosts, and amenities among them, which motivates aspect-level measurement rather than one undifferentiated sentiment score. Information also survives in the shape of the rating distribution, not only its level: Zhang and Chen (2026) found that the dispersion of Airbnb ratings, across reviewers and across rating dimensions, predicts bookings in ways the mean alone does not. A related stream connects review sentiment to prices in hedonic frameworks across U.S. markets (Lin and Yang 2024; Östh et al. 2025). I do not study prices; my outcome is the platform's own displayed signal.

Nearly all of this work is cross-sectional: it asks whether text and rating agree now. The dynamic question, whether text observed at one time predicts the movement of the displayed rating afterward, is the natural next step once ratings are known to be compressed, and it is the step I take here. I make no claim to be first to compare text with ratings, to study inconsistency between them, or to apply sentiment analysis to this setting; those literatures are established. The specific object I test is narrower: cumulative pre-snapshot text as a predictor of the subsequent change in the displayed rating, confirmed on held-out markets. To my knowledge, that object is not yet documented. Figure 1 summarizes the framing. Guest experience feeds two channels that aggregate at different speeds; the estimated association runs from the fast channel's accumulated content to the subsequent movement of the slow one, conditional on the slow channel's own level.

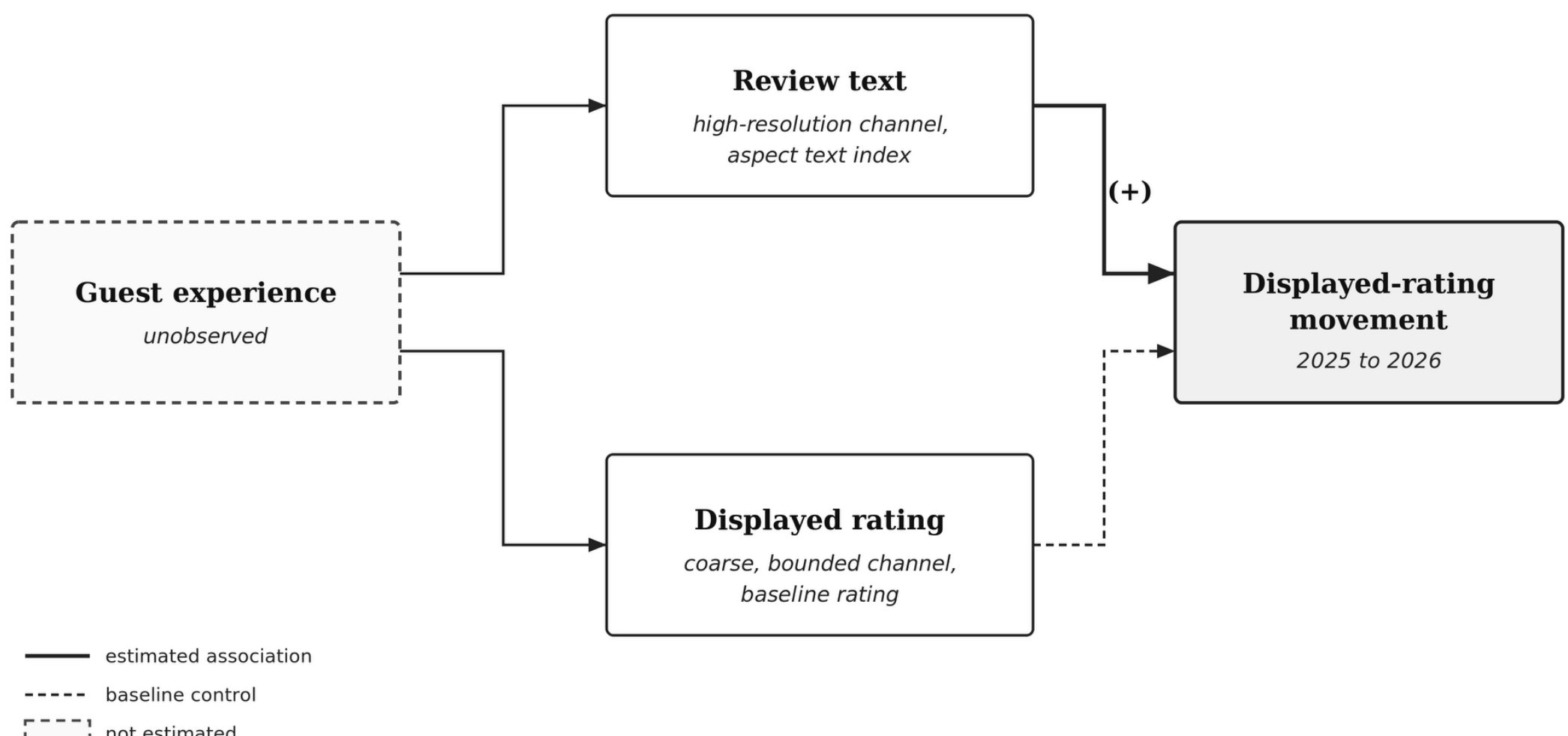

**Figure 1.** Conceptual frame: two channels aggregate guest experience at different speeds.

*Note.* The dashed construct is conceptual and is not estimated. The solid path is the estimated association; the dashed path denotes the baseline-rating control. The full control set appears in Section 4.

### 2.3 Measurement discipline for text-derived constructs

Automated text measures require validation against the construct they claim to capture (Grimmer and Stewart 2013). For sentiment specifically, van Atteveldt et al. (2021) showed that dictionary methods generally trail trained human coders and supervised models on validity, a result that disciplines what I claim. I therefore treat my index as an operational instrument, defined by its construction rule, rather than as validated human-equivalent sentiment. VADER itself was validated on social-media text against human raters at construction (Hutto and Gilbert 2014), and I add convergent and discriminant evidence in the tradition of Campbell and Fiske (1959), using the platform's own displayed aspect ratings as a second method. What I do not have, and do not claim, is a human-coded gold standard for this corpus.

## 3. Data and Measurement

### 3.1 Two snapshots of 34 markets

I collected publicly available listing and review data for 34 U.S. markets in two waves, a 2025 snapshot and a 2026 snapshot. The full-history text build processed 13,752,981 raw review rows, retained 13,748,010 reviews with usable text, split them into 50,946,917 sentences, and identified 22,429,717 aspect-sentence mentions, producing 281,086 unique listing-market rows. Matching listings across waves yields 216,248 rows present in both snapshots; 127,793 of them received at least one new review between waves, and these form the estimation panel. Every reported review count reconciles against the analytical files, and each build step wrote a manifest with row counts and checksums. I relied only on public data and collected no personal information beyond what the platform displays.

### 3.2 The aspect text index

The focal variable is a single standardized index built in five steps, all fixed before any outcome model was run. First, a prespecified dictionary assigns review sentences to four quality aspects that have displayed-rating counterparts: accuracy, cleanliness, check-in, and communication. I excluded location because it is dominated by geography and taste, and value because it evaluates price itself. Second, VADER scores each aspect-tagged sentence's sentiment (Hutto and Gilbert 2014). Third, sentence scores aggregate to a listing-by-aspect mean over the listing's complete pre-2025 review history, with a minimum of three matched text reviews per aspect. Fourth, each aspect mean shrinks toward a leave-one-listing-out market prior with empirical-Bayes strength $k = 10$, which stabilizes thin listings. Fifth, the four aspect signals standardize within market and average into one index, re-standardized within market. I refer to this variable as the aspect text index throughout, and every table uses that name.

### 3.3 Validation without human labels

Two kinds of evidence support the index as a measure of perceived quality, and one kind is absent. The convergent and discriminant evidence follows the multitrait-multimethod logic of Campbell and Fiske (1959), treating guest text and guest star ratings as two methods measuring the same four traits. Among the 83,043 listings meeting the eligibility rule, the text signal for each aspect correlates more strongly with the platform's displayed rating of the same aspect than with the displayed ratings of the other three aspects, within market, for all four aspects (Table 2). Cleanliness shows the strongest convergent correlation ($r = .63$) and check-in the weakest ($r = .38$). The separations are moderate rather than sharp, which is consistent with a broadly evaluative signal that carries some aspect-specific content, and I interpret the index accordingly as a general quality signal rather than a set of four cleanly separated aspect measures.

The absent evidence is human. No human-coded validation sample exists for this corpus, so I cannot report precision, recall, or agreement statistics against human judgment, and the dictionary literature warns that such statistics would likely be sobering (van Atteveldt et al. 2021). Every claim in this paper is therefore a claim about the defined index, not about sentiment as a human would judge it. Section 7 states the consequences.

## 4. Empirical Design

### 4.1 A locked two-wave specification

The design is a quantitative observational two-wave panel with a prespecified confirmatory protocol. Because the broader project had explored several candidate constructs on these data after the 2025 outcome data existed, an unrestricted analysis could not distinguish a real regularity from a selected one. I addressed this the way a registered analysis would. I wrote a lock file that fixed the research question, the focal variable, the estimator, the exact formula, the clustering, the falsification battery, and numeric success criteria, and I froze it before the confirmatory estimation.

The locked model regresses the change in the displayed overall rating between the 2025 and 2026 snapshots on the aspect text index. Controls are the logged number of new reviews, the logged number of lifetime reviews, market fixed effects, and fixed effects for the baseline displayed rating binned at 0.05-point width. The baseline bins matter because a listing's room to move depends mechanically on where it starts, and because regression to the mean would otherwise masquerade as a text effect. Standard errors cluster on host. Because 34 market clusters are few, the market-level p-value comes from a null-imposed Rademacher wild cluster bootstrap with 9,999 draws. I validated the bootstrap implementation against complete enumeration and an independent package before use.

### 4.2 Held-out confirmation

A seeded random split assigned the 34 markets to 17 discovery and 17 confirmation markets. The discovery half absorbed all exploratory damage. The confirmation half was touched exactly once: I estimated the locked model on it a single time, with no specification search, under three prespecified success criteria, a positive coefficient, host-clustered $p < .05$, and market wild-bootstrap $p < .05$. The lock also fixed the failure rule in advance: if any criterion failed, the finding would be reported as not independently confirmed. This protocol is a within-study replication in the spirit of a registered report, and I report the discovery and confirmation estimates separately so the reader can see both. Figure 2 summarizes the design.

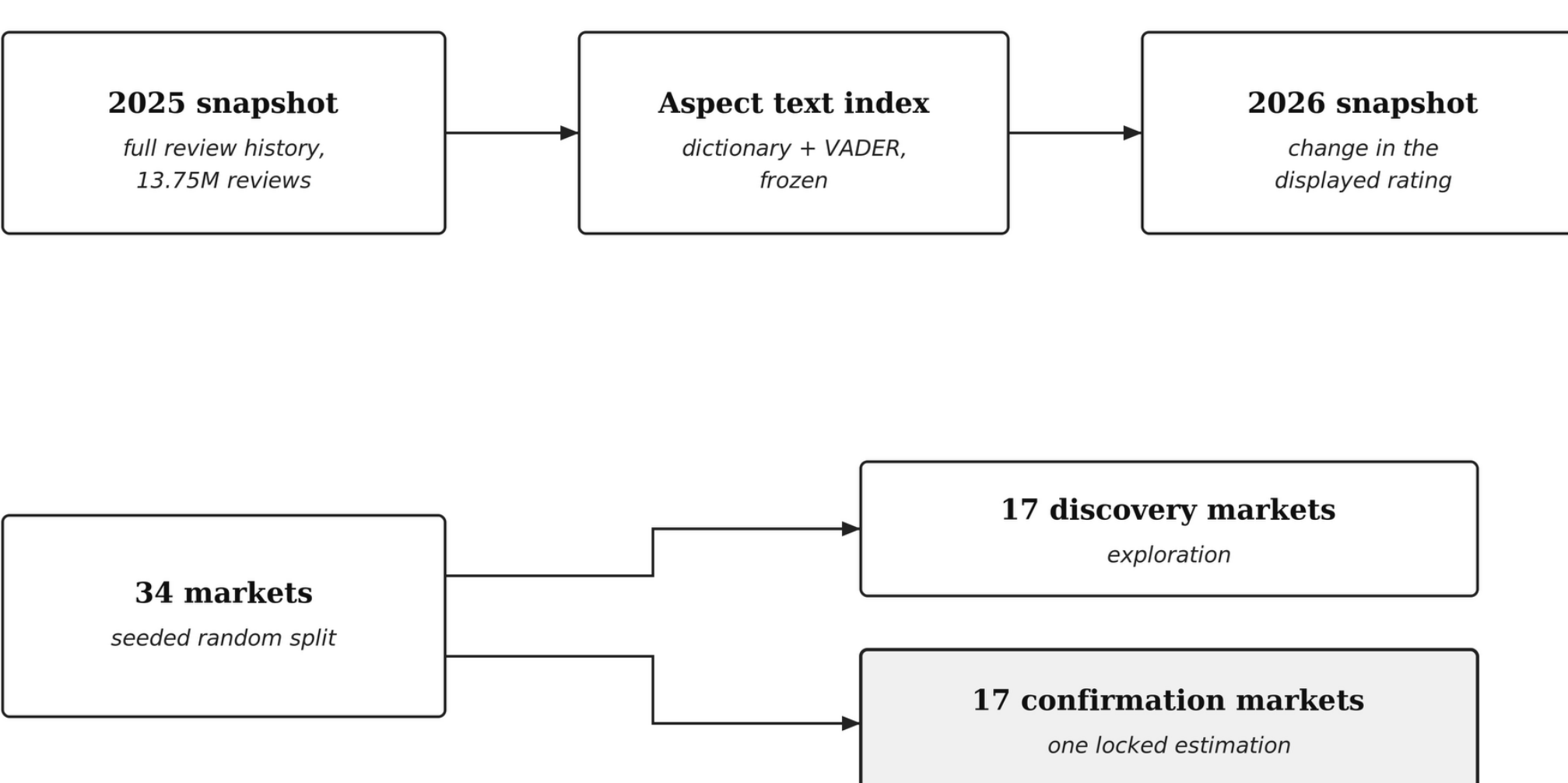

**Figure 2.** Two-wave design and the held-out-market confirmation protocol.

*Note.* Twelve months separate the snapshots. The aspect text index is computed only from reviews written before the 2025 snapshot and is frozen before any outcome model is run.

### 4.3 Falsification battery

The lock prespecified four checks. A placebo re-estimates the identical model on listings with no new reviews between waves; if the text index merely proxies for listing type or market micro-location, it would predict rating drift there too, while the causal-order logic of the design requires a null. A host fixed-effects model restricts identification to within-host variation among multi-listing hosts, absorbing everything stable about the host, including communication style and professionalism. Leave-one-market-out estimation re-runs the confirmation model dropping each market in turn. And a fine-binning check replaces the 0.05 baseline bins with 0.01 bins, tightening the regression-to-the-mean control.

## 5. Results

### 5.1 Confirmation on held-out markets

Table 1 reports the locked estimates. On the 17 confirmation markets, the aspect text index predicts the subsequent movement of the displayed rating at +0.00928 rating points per standard deviation (SE = 0.00037, host-clustered p = 5.1e-141, n = 86,349). The market wild bootstrap produced no draw as extreme as the observed statistic in 9,999 draws ($p < .0001$). The discovery-half estimate is +0.00914 (n = 41,437) and the full-sample estimate is +0.00927 (n = 127,793); the three estimates agree to the third decimal, which is what a stable regularity looks like across a random market split. All three prespecified success criteria passed.

The magnitude deserves a plain statement. A standard deviation of text warmth moves the next year's displayed rating by about one hundredth of a star. On a compressed five-star scale where most movement is small, the effect is real, precisely estimated, and modest. The finding matters because of what it implies about information flow, not because any single listing's score moves far.

### 5.2 Falsification results

Every locked check behaved as the leading-indicator interpretation requires. Among the 39,494 confirmation-half listings with no new reviews, the text coefficient is −0.0004 and indistinguishable from zero (p = .139): where no new information arrived, the old text predicts no movement. Within multi-listing hosts, with host fixed effects absorbing stable host quality, the estimate remains positive at +0.0036 (p = 3.9e-11, n = 56,198). The attenuation from +0.0093 tells me a meaningful share of the cross-sectional association reflects between-host differences. The surviving within-host component tells me it does not reduce to them. Dropping any single confirmation market leaves the coefficient between +0.0089 and +0.0096, always positive. In pre-lock exploratory runs on the full sample, tightening the baseline bins to 0.01 width left the estimate near +0.0074 with p below 1e-150. Mechanical mean reversion within coarse bins therefore does not manufacture the result. Figure 3 places the locked estimates and the falsification checks on one axis.

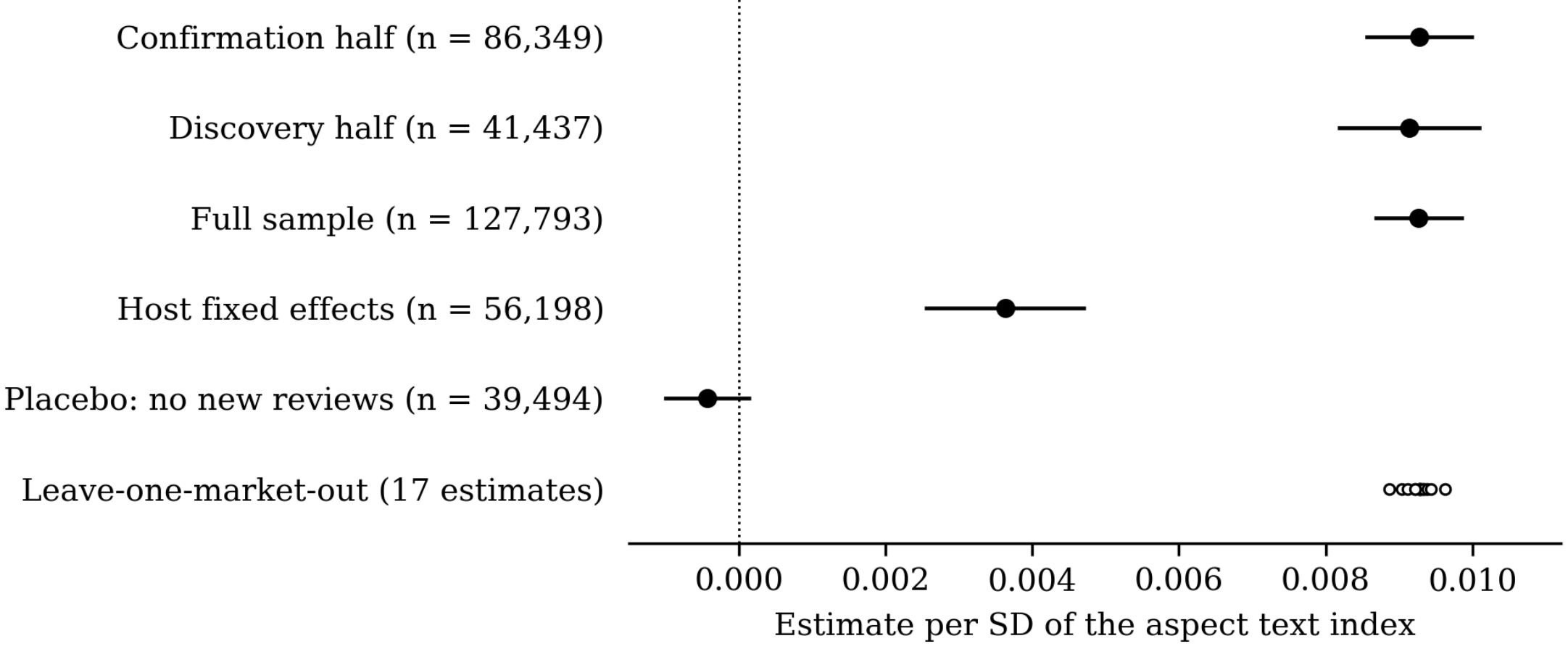

**Figure 3.** Locked estimates and falsification checks for the aspect text index.

*Note.* Points are coefficient estimates and horizontal bars are host-clustered 95% confidence intervals, from Table 1. Open circles are the 17 leave-one-market-out estimates on the confirmation half.

### 5.3 What the index measures

Table 2 reports the multitrait-multimethod correlations. Each aspect's text signal correlates more with the same aspect's displayed rating than with the other aspects' ratings: accuracy .33 versus .26, cleanliness .63 versus .46, check-in .38 versus .32, and communication .52 versus .42, all within market, n = 83,043. The pattern supports convergent validity against the platform's own second method and moderate discriminant separation. Exploratory decompositions were consistent with that moderation. The predictive signal is carried mostly by cleanliness and communication text, and it is general rather than aspect-matched. I therefore do not interpret aspect-level mismatches, and I flag any such interpretation as unsupported by these data.

**Table 1.** Locked two-wave estimates of the aspect text index on displayed-rating change.

| Model | Estimate | SE | Host-clustered p | n |
|---|---|---|---|---|
| **Confirmation half (17 held-out markets)** | +0.00928 | 0.00037 | 5.1e-141 | **86,349** |
| **Discovery half** | +0.00914 | 0.00049 | 8.3e-78 | **41,437** |
| **Full sample** | +0.00927 | 0.00030 | 1.2e-211 | **127,793** |
| **Placebo: no new reviews (confirmation)** | −0.00043 | 0.00029 | .139 | **39,494** |
| **Host fixed effects (confirmation)** | **+0.00363** | **0.00055** | **3.9e-11** | **56,198** |

*Note.* The outcome is the 2025 to 2026 change in the displayed overall rating. All models include market and baseline-rating-bin fixed effects and control for logged new and lifetime review counts. The confirmation-half market wild-bootstrap p is below .0001 (9,999 draws).

**Table 2.** Within-market correlations between aspect text signals and displayed aspect ratings.

| Text signal | Accuracy rating | Cleanliness rating | Check-in rating | Communication rating |
|---|---|---|---|---|
| **Accuracy text** | **.33** | .30 | .22 | **.26** |
| **Cleanliness text** | .55 | **.63** | .38 | **.44** |
| **Check-in text** | .33 | .29 | **.38** | **.35** |
| **Communication text** | **.44** | **.40** | **.42** | **.52** |

*Note.* n = 83,043 listings with all four displayed ratings and at least three matched text reviews per aspect. Diagonal entries are convergent correlations.

## 6. Discussion

### 6.1 Contributions

I claim two contributions, both addressed to the reputation-systems literature. First, I document that cumulative review text is a leading indicator of the displayed rating on a saturated platform, and I document it under a locked specification with held-out-market confirmation, a placebo, and host fixed effects. The saturation literature established that displayed scores separate listings poorly; this result adds that the text stream contains information the score has not yet absorbed, and that the lag is long enough to measure across a one-year window. Second, I contribute a measurement template for situations where human-coded validation is absent: define the text instrument operationally, freeze it before outcomes, and validate it against the platform's own parallel signal with multitrait-multimethod logic. The template does not replace human validation, and I do not present it as equivalent; it makes the boundary of the claim explicit instead of hiding it.

### 6.2 Mechanisms, honestly bounded

Three pathways could generate the association, and my design separates them only partially. Guests may write experiences into text with more resolution than they express in the bounded star scale, so text aggregates quality information faster than the score. Alternatively, persistent latent quality may drive both channels with the rating simply lagging, which the dominance of full-history text over recent text in exploratory runs supports. Or newly arriving guests may select listings on cues correlated with past text. The placebo rules out pathways that need no new reviews, and host fixed effects rule out stable host style as the whole story, but I cannot separate the remaining pathways with two waves, and the lock's interpretation limits say so. The claim I defend is the informational one: the displayed rating does not contain everything the platform's own text already knows.

### 6.3 Implications for platforms and hosts

For platform designers, the result identifies recoverable signal in data the platform already stores. A rating display that incorporated text-derived aspect signals would move earlier than the current aggregate, and the check-in and communication correlations in Table 2 suggest which aspects have the most uncaptured text content. For hosts, the practical reading is that text feedback anticipates where the displayed score is heading, so reading one's own review text closely is not sentimental advice but a forecast. I state these as design implications of an associational finding, not as effects an intervention has demonstrated.

## 7. Limitations

Three limitations bound the claims, and I give each its consequence. First, the text index has no human validation. Dictionary and rule-based sentiment tools measurably trail human coders on validity (van Atteveldt et al. 2021; Grimmer and Stewart 2013), so the index is best read as a reproducible instrument correlated with perceived quality, not as a measurement of it. Any construct-level conclusion inherits that uncertainty, which is why every claim in this paper is anchored to the defined index. A human-coded, multilingual validation sample remains the gate I have set for any journal submission of this work. Second, the design is a two-wave associational panel. It cannot separate faster text aggregation from latent-quality lag or from guest selection, and no causal language appears in my claims. Third, the displayed rating is platform-mediated: Airbnb's aggregation, rounding, and display rules stand between guest input and my outcome, and a change in those rules would change the measured lag. The two-snapshot window also means I observe one year of rating movement in one platform era; the regularity's stability across eras is an open question.

## 8. Conclusion

Where displayed ratings are saturated, the interesting information is in motion, not in level. Across 34 U.S. markets, the text guests wrote before a snapshot predicts how the displayed rating moves after it, under a specification locked in advance and confirmed once on held-out markets. The estimate is small, directionally consistent everywhere I looked, and robust to the checks I committed to before looking. An earlier version of this paper asked whether reviews still matter when nearly all of them are positive. The answer this version supports is that the text still carries signal; it is the display that runs behind.

## Declarations

Ethics approval and consent to participate not applicable. This study used only publicly displayed platform data and collected no personal information.

Consent for publication: not applicable.

Funding: not applicable.

Use of large language models: a large language model was used for editorial assistance with prose. I made all research design, data, analysis, and interpretation decisions, and I verified every number and reference against source files.